# The New Hard X-ray Mission

**Gianpiero Tagliaferri[1], Giovanni Pareschi**
*INAF – OABr*
*Via Bianchi 46, 23807 Merate, Italy*
*E-mail:* `gianpiero.tagliaferri@brera.inaf.it`

**Andrea Argan**
*INAF - Headquarters*
*V.le del Parco Mellini 84, 00136 Roma*
*E-mail:* `andrea.argan@inaf.it`

**Ronaldo Bellazzini**
*INFN-Pisa*
*Largo Bruno Pontecorvo 3, 56127 Pisa, Italy*
*E-mail:* `ronaldo.bellazzini@pi.infn.it`

**Enrico Costa**
*INAF – IASF_Roma*
*Via del Fosso del Cavaliere 100, 00133 Roma*
*E-mail:* `enrico.costa@iasf-roma.inaf.it`

**Osvaldo Catalano, Giancarlo Cusumano**
*INAF – IASF_Palermo*
*Via Ugo La Malfa 153, 90146 Palermo, Italy*
*E-mail:* `osvaldo.catalano@iasf-palermo.inaf.it`

**Fabrizio Fiore**
*INAF - OARm*
*Via di Frascati 33, 00040 Monteporzio Catone, Italy*
*E-mail:* `fiore@oa-roma.inaf.it`

**Carlo Fiorini**
*Università Politecnico di Milano*
*Via Ponzio 34/5, 20133 Milano, Italy*
*E-mail:* `fiorini@elet.polimi.it`

**Giuseppe Malaguti**
*INAF – IASF_Bologna*
*Via Gobetti 101, 40129 Bologna, Italy*
*E-mail:* `malaguti@iasfbo.inaf.it`

---

[1] Speaker

**Giorgio Matt, Cesare Perola**
*Università degli Studi Roma Tre*
*Via della Vasca Navale 84, 00146 Roma, Italy*
*E-mail:* `matt@fis.uniroma3.it`

**Sandro Mereghetti, Gabriele Villa**
*INAF – IASF_Milano*
*Via E. Bassini 15, 20133 Milano, Italy, Country*
*E-mail:* `sandro@iasf-milano.inaf.it`

**Giuseppina Micela**
*INAF - OAPa*
*P.zza del Parlamento 1, 90134 Palermo, Italy*
*E-mail:* `giusi@astropa.inaf.it`

The Italian New Hard X-ray Mission (NHXM) has been designed to provide a real breakthrough on a number of hot astrophysical issues that includes: black holes census, the physics of accretion, the particle acceleration mechanisms, the effects of radiative transfer in highly magnetized plasmas and strong gravitational fields. NHXM is an evolution of the HEXIT-Sat concept and it combines fine imaging capability up to 80 keV, today available only at $E<10$ keV, with sensitive photoelectric imaging polarimetry. It consists of four identical mirrors, with a 10 m focal length, achieved after launch by means of a deployable structure. Three of the four telescopes will have at their focus identical spectral-imaging cameras, while X-ray imaging polarimetric cameras will be placed at the focus of the fourth. In order to ensure a low and stable background, NHXM will be placed in a low Earth equatorial orbit. In this paper we provide an overall description of this mission that is currently in phase B.

# 1. Introduction

The exploration of the X-ray sky during the 70s–90s has established X-ray astronomy as a fundamental field of astrophysics. However, the emission from astrophysical sources is by large best known at energies below 10 keV. The main reason for this situation is purely technical since grazing incidence reflection has so far been limited to the soft X-ray band. Above 10 keV, i.e. in the so called hard X-ray band, so far we could use only passive collimators, as those on board of the *Beppo*SAX and *Rossi*XTE satellites, or coded masks as those on board of the INTEGRAL and Swift satellites. As a result less than two thousands sources are known in the whole sky in the 10–100 keV band, a situation recalling the pre–*Einstein* era at lower energies. Yet these results demonstrated the fundamental importance of the wide band to understand in depth the physics of the variety of objects in the X-ray sky and their emission and transfer processes. These include, for instance, thermal versus non-thermal processes, particle acceleration, radiative transfer in strongly magnetized media, jet formation and evolution, absorption and scattering in Compton thick surroundings: all of these in often (highly) variable sources. Those results demand that above 10 keV, in order to match the sensitivity below 10 keV, one should go down to limits in flux three orders of magnitude better than those reached so far, and correspondingly out to very large cosmological distances. This last point brings to the forefront the topical case of the Cosmic X-Ray Background (CXB). Made of discrete sources in the soft X-rays, consisting mainly of accreting Super-massive Black Holes distributed over cosmological distances, in the spectral region where it peaks, namely between 20 and 30 keV, a fraction as large as about 98% of the CXB remains unresolved. Therefore the putative discrete sources that, according to model-dependent extrapolations, should be contributing to it are still to be identified. All this calls for focusing X ray optics at higher energies, up to 80 keV and beyond.

Polarimetry is another field in which X-ray astronomy has not progressed much. Despite the fact that linear polarization, in different degrees according to the electromagnetic process at work, is expected to be a distinctive feature of the X-ray emission in quite a variety of object classes, and therefore a clue, in some cases the unique clue, to the correct physical model, the lack of an efficient measurement tool has so far restricted to spectra and variability the parameters on which the models are built. To date X-ray polarization has been detected only in a very bright and highly polarized source, the Crab Nebula observed "as a whole", that is without resolving its complex structure. The advent of a new generation of detectors [1], when combined with large area X-ray telescopes, can now provide a huge increase in sensitivity, putting polarimetry at pace with the other fields of X-ray astronomy.

On the basis of the above considerations, the New Hard X-ray Mission (NHXM) is currently under development in Italy and it is meant to provide a real breakthrough on a number of hot astrophysical issues, by exploiting the most advanced technology in grazing incidence mirrors, in spectro-imaging focal plane cameras and in polarimeters. The NHXM is based on four identical telescopes with a focal length of 10 m, achieved after launch by means of a deployable structure (see Fig. 1). The four identical mirror modules (MM) will, for the first time, extend from 0.5 keV up to 80 keV the fine imaging capability today available only at E<10 keV. At the focus of three telescopes there will be three identical spectro-imaging cameras, at the focus of the fourth there will be two polarimetric cameras covering the range 2-35 keV. For the first time it will be possible to obtain simultaneously images, spectral energy distributions and polarization properties (degree and angle) of sources in the X-ray sky over such a large energy range.

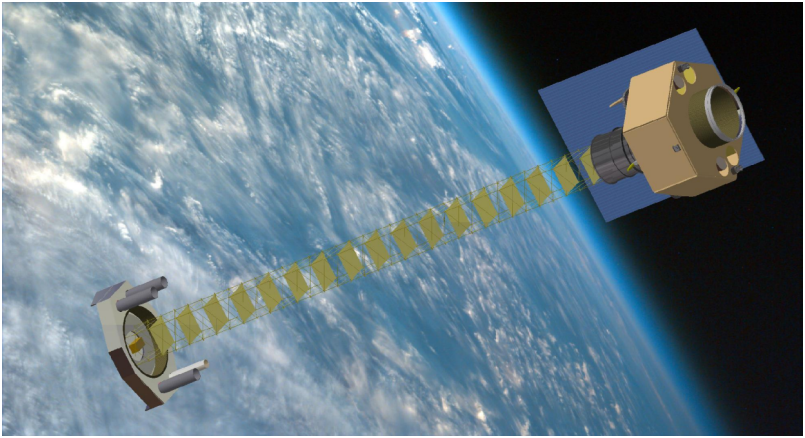

**Figure 1:** artistic view of the NHXM satellite. The four mirror modules are on the satellite platform and the focal plane cameras on the detector platform at the end of the deployable truss.

The NHXM is being developed in Italy as an evolution of the original HEXIT-Sat project, improved by the addition of imaging polarimetric capabilities. HEXIT-SAT has been studied together with Simbol-X as part of an industrial Phase-A contract commissioned to Thales Alenia Space Italy, with the participation of the Media Lario Company and of the INAF institution. After that the French-Italian-German Simbol-X mission [2] has been cancelled due to programmatic problems, the NHXM became the reference project for the Italian high energy community. We will now describe the main characteristics of the NHXM mission.

## 2. The mission configuration

NHXM will operate in a circular, low inclination orbit at 600-km mean altitude, that provides a very low background, as proven by the BeppoSAX and Swift missions,. The satellite will be stabilized on three axes with good pointing capability. It will host 4 X-ray telescopes with at their focus either a focal plane camera (for three of them) hosting a combination of two detectors sensitive from 0.5 up to 80 keV or two interchangeable polarimeter cameras covering the range 2-35 keV at the fourth telescope. The four MM will achieve a net area comparable or higher to that one of the Simbol-X single MM (see next section), with the following advantages: *i)* each MM is smaller and lighter; *ii)* segmentation in four telescopes allows to implement a redundancy policy; *iii)* it is possible to implement polarimetric cameras in one of the telescopes.

The NHXM satellite has been designed to accommodate two sets of payload elements (optics and focal planes) on different modules, separated by 10 m. This distance will be maintained with a given alignment and stability. Moreover, the whole system will be compatible with VEGA, the smallest European launcher. The launch is foreseen in 2016 for a mission lifetime of 3 (+2, goal) years, with Malindi used as the ground station. NHXM will be operated as a X-ray observatory. The service platform is derived from the service module already used for the BeppoSAX mission and will make an extensive reuse of the subsystems of PRIMA. In particular, the service module will carry aboard the following subsystems: *i)* the central cylinder (main platform structure) hosting the extendible bench canister; *ii)* the extendible bench connecting the extendable platform, that accommodates the Detector Modules; *iii)* the four MM, arranged around the external wall of the central cylinder.

The platform cylinder and the focal plane assembly (FPA) are arranged in the Detector Platform located at the focal length of 10m by the deployable truss (Fig. 1). During the NHXM phase A study a trade-off considering several extensible boom concepts (inflatable structure, telescopic boom, coilable truss, circulated truss) has been performed. This trade-off allowed the identification of the articulated truss boom for the mission implementation. The material used in this kind of structure is a carbon fibre reinforced polymer. After injection into orbit and commissioning, the expandable truss will position the Detector Platform at the operating mirror focal lengths of 10 m. The Detector Platform provides also the thermal control of the FPA, the mechanical structure, the harness connecting the Detector equipment and the metrology devices. The driving performance requirements of this architecture concerns is the focal plane stability over the timescale of an observation, and under large temperature gradients (both axial and circumferential). The focal length must not change by more than a few mm and the alignment between the optics reference axis and the focal plane reference axis must be stable within 1.5 arcmin. This condition is mainly guaranteed by the selection of the material and by the thermal control of the Detector Platform and expandable truss. The residual variations of the focal plane position (impacting on HEW) are monitored by a Service Metrology system allowing for a post-facto image reconstruction.

### 3. The scientific payload configuration

We have a very stringent and demanding set of scientific requirements that must be satisfied and that are very similar to those of Simbol-X [2], plus the polarimetric requirements. In Tab. 1 below we report the most relevant ones. We describe now the scientific payload.

**The mirror modules:** the NHXM mirrors will be electroformed nickel-cobalt alloy (NiCo) shells with Wolter I profile. The adopted technology has been successfully used for the gold coated X-ray mirrors of Beppo-SAX, XMM-Newton and Jet-X/Swift satellites (although for these mission the mirror shell were made in pure nickel, while for NHXM they will be made in NiCo that has better stiffness and yield properties). This technology has been developed and consolidated in the past two decades in Italy by the INAF Brera Astronomical Observatory in collaboration with the Media Lario Technology Company. For the NHXM mirrors, a few important modifications are foreseen: 1) the use of multilayer reflecting coatings, allowing us to

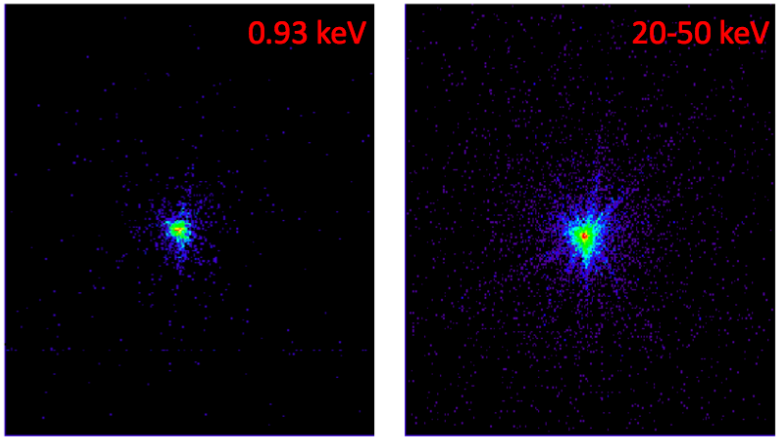

**Figure 2:** X-ray point source image obtained at the Panter-MPE facility for a prototype with two mirror shells: *(left)* HEW=18 arcsec at 1 keV;

*(right)* HEW=25-30 arcsec in the band 20-50 keV (see also [3]).

obtain a larger FOV and an operative range up to 80 keV and beyond; 2) the NiCo walls will be a factor of two thinner than the XMM Ni-mirror shells, to maintain the weight as low as possible. With respect to the first point, once the gold-coated NiCo mirror shell has been replicated from the mandrel, the multilayer film will be sputtered on the internal surface of the shell by using a two-targets linear DC magnetron sputtering system. This process has been developed and tested for monolithic pseudo cylindrical shells at Media Lario, where a multilayer coating facility has been developed and installed as part of the Phase A activities.

**Table 1:** some of the NHXM top-level scientific requirements.

| Parameter | Value |
| --- | --- |
| Energy band: | 0.5 – 80 keV |
| Field of view (at 30 keV): | ≥ 12' (diameter) |
| On-axis effective area: | ≥ 100 $cm^2$ at 0.5 keV;     ≥ 1000 $cm^2$ at 2-5 keV<br>≥ 600 $cm^2$ at 8 keV;    ≥ 350 $cm^2$ at 30 keV<br>≥ 100 $cm^2$ at 70 keV;    ≥ 50 $cm^2$ at 80 keV (goal) |
| Detectors background | < 2×$10^{-4}$ cts $s^{-1}cm^{-2}keV^{-1}$ HED<br>< 2×$10^{-4}$ cts $s^{-1}cm^{-2}keV^{-1}$ LED |
| Angular resolution | ≤ 20"*(HPD)*, E < 30 keV; ≤ 40"*(HPD)* at E = 60 keV (goal) |
| Polarisation sensitivity | 10% MDP for 1 mCrab source in the 2-10 keV band and for a 3 mCrab source in the 6-35 keV band, in 100 ks |

Engineering Models (EM) with two integrated shells have already been developed and tested at the Panter-MPE X-ray calibration facility [3]. Figure 2 shows two images taken at 0.93 keV and in the band 20-50 keV. Note how the image quality is extremely good also at the higher energies. The structure of the spider arms can be seen. The HEW at 0.93 keV is about 18 arcsec, while that one at 30 keV is ~25 arcsec, not very far away from the requirements (see Tab. 1). This somewhat higher value is due to a higher mirror surface roughness, with respect to the one required by the mission. This aspect is now under investigation as part of the new technological development program that ASI has assigned to Media Lario. As a result a new EM with three shells has been developed and will soon be tested at the Panter facility.

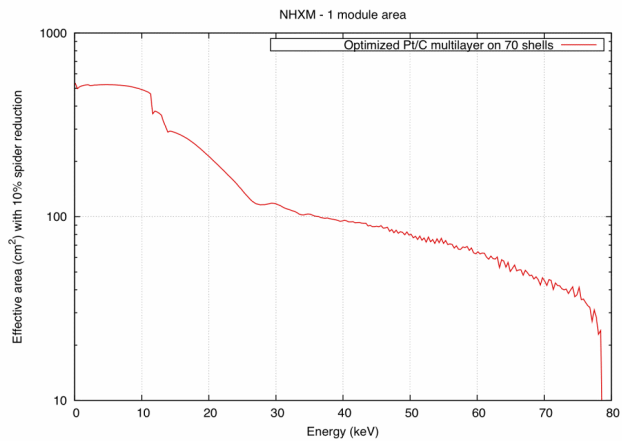

**Figure 3:** effective area of one single NHXM mirror module

For Simbol-X the optical design was achieved with a single large-aperture MM with a very long focal length of 20 m [4]. A longer focal length ensures that shells with larger diameter will efficiently reflect high-energy photons. If we decrease the focal length then only smaller diameter shell will efficiently focus high-energy photons, implying a smaller effective area. Therefore, to reach the previous values we need to add more MM. There are advantages and disadvantages in this solution of course. From the scientific side a big advantage is that we can add polarimetric capability to one of the telescopes. From the mirrors point of view the advantage is that we can replicate more shells from the same mandrels, minimising the number of mandrels and making them smaller. Being the fabrication of the mandrels one of the most critical items in the mirror production, these are both big advantages. From the detector point of view, the advantage is redundancy, if one detector fails we can recover this by simply extending the exposure time. The disadvantage being that we need to develop more than one detector of course. The current baseline of the NHXM optical design foresee 70 shells for each MM with a diameter ranging from 15.5 to 39 cm. The effective area of a single MM is reported in Fig. 3.

**The focal plane spectro-imaging and polarimetric camera**: three spectro-imaging cameras have to match the high-energy X-ray multilayer optics performance over the broad 0.5 - 80 keV energy range, providing a very low background environment. A hybrid detector systems will be used with two detection layers, plus an effective anticoincidence system:

- The Low Energy Detector (LED), with the role to detect with a high quantum efficiency and energy resolution the soft X-ray photons between 0.5 and 15 keV, placed on the top of the focal plane assembly. The baseline configuration foresees an Active Pixel Sensor (APS) based on DEPFET readout, developed by MPE [5] in collaboration with the Department of Electronics of the Politecnico of Milano for the implementation of the fast read-out system VELA [6] (see Fig. 4), a solution already assumed for the former SIMBOL-X mission;

- The High Energy Detector (HED), placed in series beyond the LED, with the role to detect with a high quantum efficiency and energy resolution the hard X-ray photons in the 10-80 keV band. It will be formed by four pixellated, 1mm thick and 2x2 cm size each, Schottky barrier CdTe crystals. Each pixel is connected to its own read-out ASIC electronics by a proper bonding. A first prototype using the same ASIC developed for the polarimeter [XPOL, e.g. 7] as already been made, while a new version with the required pixel size (250 µm instead of the current 80 µm) is under development;
- The Anti-Coincidence system (AC), completely surrounding the LED and HED systems, in order to efficiently screen the particle and gamma ray background. The AC will be realized using well-known inorganic scintillators like NaI or CsI crystals, already successfully used in a number of focal plane detectors aboard X- and gamma-ray telescopes. As an alternative, the use of new scintillators as e.g. LaBr3, currently investigated by ESA, is very attractive due to the fast rise time and high light output.

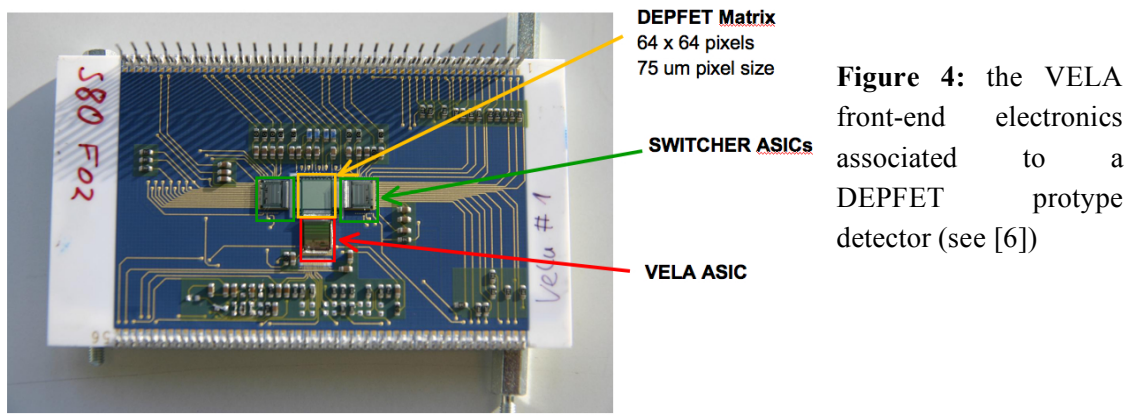

**Figure 4:** the VELA front-end electronics associated to a DEPFET protype detector (see [6])

Also for the polarimetric camera, at the focal plane of the fourth telescope, the goal is to cover an energy range as large as possible. The purpose of the focal plane polarimeter is to provide polarization measurements simultaneously with angular, spectral and timing (at few ms level) measurements. The instrument is based on a Gas Pixel Detector, a position-sensitive counter with proportional multiplication and a finely subdivision of the charge collecting electrode in such a way that photoelectron tracks can be accurately reconstructed and their emission direction derived [1,7]. The linear polarization is determined from the angular distribution of the photoelectron tracks. The Gas Pixel Detector is based on a gas cell with a thin entrance window, a drift gap, a charge amplification stage and a multi-anode readout plane, which is the pixellated top metal layer of a CMOS ASIC analog chip. The polarimeter already developed by the INFN of Pisa and INAF-IASF of Roma (see Fig. 5) currently covers the band from 1 to 10 keV. By changing the gas pressure and mixture inside the detector cell it is possible to get polarization measurements up to 30 keV. However, to cover this larger band, it is necessary to use two detectors that can be put at the focus of the telescope by means of a rotating wheel.

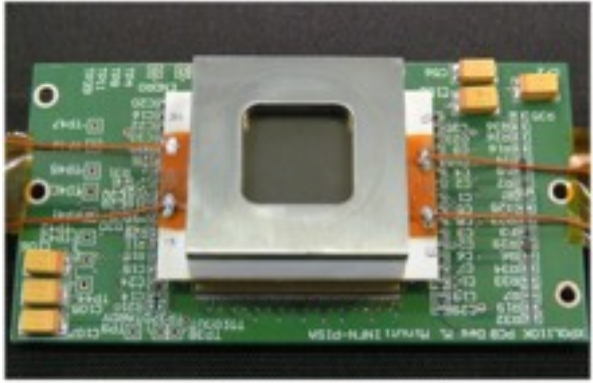

**Figure 5:** the sealed Gas Pixel Polarimetric Detector with the 50 nm Be window [7].

## 4. Conclusions

The NHXM is being developed in Italy as an evolution of the original HEXIT-SAT project, improved by the addition of imaging polarimetric capabilities. There are currently two contracts financed by the Italian Space Agency; one with the Media Lario Technology company for the development of the X-ray optics foreseen for this mission and one with INAF, that foresee a number of activities at different Institutes of INAF, INFN and Universities. These activities include the definition and finalisation of the scientific requirements of the mission, but also the development of prototypes for the scientific payload to consolidate their feasibility. Moreover, one of the goals is also to look for possible strong international collaborations.
**Acknowledgements:** we acknowledge the support of the grant ASI/INAF I/069/09/0.